# Efficient room-temperature light-emitters based on partly amorphised Ge quantum dots in crystalline Si

M. Grydlik*, F. Hackl, H. Groiss, M. Glaser, A. Halilovic, T. Fromherz, W. Jantsch, F. Schäffler and M. Brehm*

*Institute of Semiconductor and Solid State Physics, Johannes Kepler University Linz, Altenbergerstrasse 69, A-4040 Linz, Austria*

*email: moritz.brehm@jku.at
        martyna.grydlik@jku.at

**INTRODUCTORY PARAGRAPH**

**Semiconductor light emitters compatible with standard Si integration technology (*SIT*) are of particular interest for overcoming limitations in the operating speed of microelectronic devices[1-3]. Light sources based on group-IV elements would be SIT compatible but suffer from the poor optoelectronic properties of bulk Si and Ge. Here, we demonstrate that epitaxially grown Ge quantum dots (QDs) in a fully coherent Si matrix show extraordinary optical properties if partially amorphised by Ge-ion bombardment (GIB). The GIB-QDs exhibit a quasi-direct-band gap and show, in contrast to conventional SiGe nanostructures, *almost no thermal quenching* of the photoluminescence (PL) up to room-temperature (RT). Microdisk resonators with embedded GIB-QDs exhibit threshold-behaviour and super-linear increase of the integrated PL-intensity ($I_{PL}$) with increasing excitation power $P_{exc}$ which indicates light amplification by stimulated emission in a fully SIT-compatible group-IV nano-system.**

**MAIN TEXT**

Silicon micro- and nanophotonics is a field of tremendous research interest: it connects Si-photonics with Si-based microelectronics and aims e.g. to improve on-chip data communication and processing by using guided light for data transfer instead of copper wires[1-3].

Here, the main problem arises from the poor light-emission properties of Si and Ge that result from the indirect bandgap of crystalline group-IV materials. Recent breakthroughs in Si-photonics include demonstrations of a highly doped, electrically pumped Ge-laser[4], group-III-V QD-lasers bonded onto Si substrates[5], group- III-V QD-lasers epitaxially grown on Si substrates[6] and PL from strained Ge[7,8]. Most recently, a SnGe-laser grown on a Si substrate was demonstrated to operate up to 90K[9] but shows strong thermal quenching of the SnGe-PL (about two orders of magnitude) between 20 K and 300K[9].

Utilizing group-IV nanostructures as Si-based light-emitters would allow omitting the growth of thick, dislocation-rich GaAs or SiGe graded buffer layers, necessary for some of the approaches described above. The development of epitaxially grown, strain-driven, coherent Ge QDs on Si (Ge/Si-QDs)[10,11] gave hope that quantum-confinement effects in group-IV nanostructures could be used to produce high-performance optical devices. But the optical properties never matched expectations. Because of the relatively low lattice-mismatch between Si and Ge (~4%), the resulting QDs become at least in one spatial direction larger than several tens of nm. Their carrier wave-functions have one- or two-dimensional (1D, 2D) character rather than the zero-dimensional (0D) aimed at. Another drawback is the spatially indirect recombination path in Ge/Si-QDs due to a type-II band-alignment[12], where only holes are confined in the Ge-QDs.

Because of strong quantum confinement in all three spatial dimensions (3D), porous silicon and Si nanocrystals (Si-NCs) show significantly better optical properties at RT[13,14,15]. Although these materials are very promising, no RT continuous-wave lasing was reported so far.

In this work, we close the gap between SIT-compatible Ge/Si-QDs[16] and porous Si/ Si-NCs with their superior PL-properties.  For this purpose, we bombarded self-organized Ge/Si-QDs during epitaxial growth with Ge-ions. This leads to partially amorphised Ge-QDs embedded in a dislocation-free Si matrix. The strain-driven Ge/Si-QD-formation occurs on a supersaturated Ge wetting layer (WL), once a critical thickness of ~4.2 monolayers (~0.6nm) is exceeded[17]. For a deposition of 0.7nm of Ge at 500°C small hut-shaped QDs[11] form with a height of about 3 nm. The dot-density can be tuned between about $2 \cdot 10^{10}$ cm$^{-2}$ (Fig. 1a) and ~$2 \cdot 10^{11}$ cm$^{-2}$ (Fig. 1b) by varying the Ge coverage. During Ge deposition, the sample is bombarded by positively charged Ge ions (dose: ~$10^4$ μm$^{-2}$) that are accelerated by voltages $V_{GIB}$ of down to -2.8 kV. The high-resolution transmission electron microscopy (TEM) image in Fig. 1c reveals a partly crystalline and partly amorphised GIB-QD embedded in a crystalline Si-matrix. In the TEM images, the amorphisation results in a blurring of the {111}-crystal-lattice fringes due to distorted atom-positions, in contrast to the regular atomic arrangement in the Si layer below and above the GIB-QD. The former hut-cluster is thus divided into small crystalline Ge regions with diameters below ~4 nm and glassy Ge regions with higher energy bandgap[18]. Si, due to its higher displacement-energy[19], is less likely to be disrupted upon GIB-treatment. No amorphisation of the Ge-WL is observed (see supplementary material).

In Figure 2 we discuss the main PL-features of the GIB-QDs. In Fig. 2a we show excitation-power ($P_{exc}$) dependent PL-spectra, normalized to the intensity maximum of the GIB-QD-PL and recorded for a sample temperature $T_{PL}$ of 6 K. With increasing $P_{exc}$ from 1.5 µW to 1600 µW the onset of the PL shifts from about 1450 nm to 1250 nm, while the intensity-maximum of the GIB-QD-PL moves from 1560 nm to 1340 nm. The PL-spectra were fitted using four Gaussians with maxima at (i) 1552 nm, (ii) 1488 nm, (iii) 1420 nm and (iv) 1352 nm and a common full-width at half-maximum of 70 nm, as shown in the inset of Fig. 2b for $P_{exc}$ = 1070 µW. In Fig. 2b, the integrated PL intensity, $I_{PL}$, of the GIB-QDs as well as the sum of the two Gaussians with longer and shorter wavelength (i and ii) and (iii and iv), respectively, are plotted versus $P_{exc}$. The PL-contributions at longer wavelength (i and ii) follow a power-law of $m \approx 0.6$ according to $I_{PL}(P_{exc}) \sim P_{exc}{}^{m}$, while those at shorter wavelengths (iii and iv) yield $m \approx 1$. In Fig. 2c, we present PL-spectra ($P_{exc}$ = 1600 µW) of GIB-QDs, obtained for $T_{PL}$ of 20 K, 80 K, 197 K and 300 K. The spectra are *to scale*, i.e., the PL-intensity hardly decreases in this temperature range. Furthermore, at higher $T_{PL}$ the onset of the GIB-QD-PL shifts to about 1200 nm. The temperature-dependent behaviour of the GIB-PL for $P_{exc}$ of 1600 µW, 430 µW and 136 µW is depicted in the Arrhenius plots in Fig. 2d. In the inset the data are plotted on a double logarithmic scale to emphasize the PL-quenching at high $T_{PL}$. The solid red lines are fits to the data corresponding to activation energies $E_A$ of ~350 meV for both, $P_{exc}$=136 µW and 430 µW (see supplementary material).

We attribute the pronounced shift of the GIB-QD-PL to shorter wavelengths for increasing $P_{exc}$ to a combination of i) progressive filling of smaller GIB-QDs that, due to higher confinement-energies, exhibit higher PL-transition energies and ii) state-filling in the GIB-QD potentials. Thus, for very low $P_{exc}$, only ground-states of the largest GIB-QDs are filled. For epitaxial Ge/Si-QDs $I_{PL}$ usually increases with $P_{exc}$, following a sub-linear power law of about $m$=0.6 caused by Auger-recombination[20]. A power law with $m$=1, as found here for the shorter wavelength part of the GIB-QD-PL spectrum, usually occurs for direct-gap semiconductor QDs, e.g. in the III-V material system[21]. Thermal quenching was reported for epitaxial Ge/Si-QDs with $E_A$ ~60 meV[20]. This is in sharp contrast to the much higher $E_A \approx 350$ meV which we extracted here for the GIB-QDs. We tentatively ascribe this high $E_A$ to electron-occupancy of trivalent Ge-bonds in the glassy region of GIB-QDs (see discussion of Fig. 5), in analogy to similar deep levels (~250-350 meV) in Si[22]. From the discussion of Fig. 2 we conclude that small GIB-QDs are quasi-direct bandgap semiconductor nanostructures. This conclusion is supported by (i) the observed power law of $m$=1, (ii) the high $E_A$ in combination with the temperature-stability of the GIB-QD-PL, and (iii) the fact that the k-selection rule of the indirect band-gap will be softened due to Heisenberg's uncertainty principle because of electron localization at the trivalent Ge bonds and highly localized hole-states in the GIB-QDs with diameters < ~4 nm. This interpretation is further supported (iv) by the strong decrease of the average PL-decay lifetime $\tau_{av}$ of the GIB-QDs with decreasing GIB-PL-emission wavelength $\lambda_{PL}$. At low $T_{PL} \leq 45$ K, $\tau_{av}$ decreases from about 500 ns for $\lambda_{PL}$=1565 nm to about 600 ps for 1330 nm (supplementary material). The latter PL-decay time is significantly faster than the ~20 ns observed for Ge/Si-QD-multilayers[23]. The short $\tau_{av}$ is also consistent with the fact that the PL-signal does not saturate even at high $P_{exc}$ (Fig. 2b), and that radiative recombination can compete with Auger-recombination. When thermal quenching of the PL is observed at very high $T_{PL}$, $\tau_{av}$ approaches *independently* of $\lambda_{PL}$ a value between 1 and 2 ns, caused by non-radiative recombination processes (see supplementary material).

One remaining question certainly regards the microscopic origin of the GIB-QD-PL. In the supplementary material we provide evidence for the hole-ground-states being confined within the crystalline parts of the GIB-QDs. With increasing $V_{GIB}$, or with increasing implantation dose, the volume of the glassy regions in the GIB-QDs increases. Consequently, the confinement shift of the hole-states shifts the GIB-QD-PL to shorter wavelengths.

To further explore the prospects of GIB-QDs as a promising material for SIT-compatible light sources, we incorporated them into a microdisk resonator[24] of 1.8 μm diameter. The excited resonant modes are whispering-gallery-modes (WGM) running around the disk's circumference, and radial Fabry-Pérot-modes (FPM) across the disks (see inset in Fig. 3). The former appear as sharp emission lines, the latter as broader peaks. Figure 3 displays PL-spectra of a microdisk for $T_{PL}$=10 K and RT and PL-spectra for increasing $P_{exc}$ at $T_{PL}$=10 K. In the following we will focus on the emergence of the Lorentzian–shaped transversal electrical mode TE(12,1), emitting at about 1323 nm. In Fig. 4 its integrated PL intensity, $I_{WGM}$ is plotted versus $P_{exc}$ on a double-logarithmic scale, and, in the inset, on a double-linear scale. The $P_{exc}$-dependence displays threshold-behaviour at about 100 μW. For higher $P_{exc}$, $I_{WGM}$ tends to saturate to a value of $m$=1 (Fig. 4). Such a threshold-behaviour and s-shaped $I_{PL}$-curves are, albeit less pronounced, still present up to RT and are indicative of stimulated emission. However, due to the strong filling effects of smaller GIB-QDs with higher $P_{exc}$, a natural threshold-behaviour of a cavity-mode emitting at 1323 nm is predicted (see Fig. 2a). To unfold the influence of the GIB-QD-filling on the threshold-behaviour of the TE(12,1) mode, we plotted in the inset of Fig. 4 the ratio $I_{WGM}/I_{FPM}$ versus $P_{exc}$. Here $I_{FPM}$ stands for the $I_{PL}$ of the FPM and both $I_{WGM}$ and $I_{FPM}$ emit at 1323 nm but couple to different GIB-QDs. Thus, if the threshold-behaviour would be caused only by the $P_{exc}$-driven filling of the GIB-QDs we would expect $I_{WGM}/I_{FPM}$ to remain constant vs. $P_{exc}$. However, at the threshold a distinct increase of $I_{WGM}/I_{FPM}$ from 3 to about 11 is observed. Also, we observe linewidth-narrowing of the emission-mode which is consistent with the threshold for stimulated emission (supplementary information).

Finally, we propose mechanisms for the formation of GIB-QDs and their overgrowth with crystalline Si (Fig. 5). When a low-energy Ge-ion (≤3keV) impinges on the sample, it undergoes collisions with the atoms near the surface causing their displacement since the ion-energy is significantly larger than the displacement-energy of Si and Ge (~14eV)[19]. If these recoil atoms obtain more energy than the displacement-energy they themselves can displace further atoms creating a collision-cascade of ellipsoidal shape[25] (Fig. 5a) until the energies of ion and recoil atoms are dissipated. The cascade of a single Ge-ion (≤3keV) produces one small, a few nm wide amorphous zone[25]. During Ge-deposition and ramp-down to the growth-temperature of the Si capping-layer, solid-phase epitaxial regrowth (SPER) takes place, in which amorphised Ge is recrystallised if the sample-temperature is above about half of the melting temperature of the crystal[26]. Recrystallization occurs because crystalline material is in a lower energy configuration than amorphous one due to the lack of bond-angle-distortions[27]. SPER also uses kink-sites[28] of the still crystalline surface to laterally recrystallize the localised amorphous zones (Fig. 5b) on top of which later the crystalline Si matrix can be grown (Fig. 5c). In Fig. 5d we present a band-alignment scheme that is consistent with the observed $\lambda_{PL}$, the high $E_A$, a power-coefficient of $m$=1, and the negligible thermal-quenching of $I_{PL}$ at RT. Electrons tunnel to dangling-bond-states in the amorphous Ge regions from where they recombine spatially indirect with hole-states located in the crystalline Ge (long $\lambda_{PL}$, type-II transition, red arrow in Fig. 5d) and with hole-states with wave-function-maxima in the amorphous Ge (short $\lambda_{PL}$, type-I transition, blue arrow in Fig. 5d). Excitation of an ensemble of GIB-QDs with different sizes of the glassy and crystalline regions leads to the observed broad range of transition energies. Thermal quenching will be observed if either holes escape from the GIB-QDs, or electrons localized at dangling bonds overcome an $E_A$ of about 300meV.

In summary, we expect GIB-QDs to bridge the gap between epitaxial group-IV QDs and QDs in porous silicon or Si-NC-systems, and will open new paths for Si-photonics based on group-IV nanostructures. GIB-QDs, separated glassy and crystalline Ge regions in the QD, are small enough in all three dimensions to exhibit 3D-quantum-confinement, similar to that of porous Si and Si-NCs. And, just as epitaxial Ge/Si-QDs, GIB-QDs can be embedded into a fully crystalline, electrically conducting Si-matrix, making the system fully compatible with integrated Si-technology and electrical pumping.

**METHODS**

**Sample growth.** Although in this work the samples were grown by molecular beam epitaxy (MBE) in combination with in-situ low energy Ge-ion-bombardment, all fabrication steps can be in principle performed with *SIT*-compatible fabrication methods. Ge quantum dots can be grown by CVD[29] and ion-implantation and annealing are standard procedures in Si-based integrated technology.

Here, the growth of GIB-QDs was carried out in a Riber SIVA45 solid-source MBE chamber. We use buried-oxide SOI substrates with a $SiO_2$ thickness of 2 μm and a high-quality Si(001) top layer with 160 nm thickness. The samples were, after ex-situ sample cleaning, in-situ oxide-desorbed at 950°C for 15 min. Thereafter, a 40 nm thick Si buffer layer was grown at a temperature that was ramped-down from 550°C to 500°C. A single Ge layer was grown at a growth temperature of 500°C with coverages of either 7 Å or 8.4 Å. QDs grown at such low temperatures are referred to as hut cluster or huts[11]. They are confined by 11.3° steep {105}-facets, which, due to kinetic reasons[11], are elongated with a rectangular base, and thus resemble huts (see Fig. 1a-b).

During growth a small fraction of the Ge atoms evaporating from the source is ionised as they pass the electron beam of the evaporator. Those Ge ions (dose~$10^4 μm^{-2}$, i.e. about one ion per area of $10 \times 10$ nm$^2$) are then accelerated towards the substrate that is biased by an adjustable substrate bias $V_{GIB}$ between 0 and -2.8keV.

Finally, the GIB-QDs were embedded into a Si matrix by capping with Si. For this purpose, the substrate temperature is ramped-down from the 500°C used for Ge deposition to 350°C for Si-cap deposition to preserve the shape and composition of the QDs and the WL[30]. During the ramp-down the topmost part of the GIB-QDs recrystallises laterally via solid-phase epitaxial regrowth, which then allows for overgrowth with a fully crystalline Si (see Fig. 1c and for the WL in the supplementary material).

**Optical and structural investigations.** The surface topography was analysed by means of atomic force microscopy (AFM) using a Digital Instruments Dimension 3100 AFM. In order to get insight into the structural properties of the GIB dots, we performed cross-sectional transmission electron microscopy using a JEOL JEM-2011 FasTEM instrument operated at 200kV. The cross-sectional lamellae were cut by a focused ion beam using a ZEISS 1540XB CrossBeam facility.

For the PL experiments, we used an excitation diode laser operated at 442 nm and a microscope objective with a numerical aperture of 0.7 which is used both for laser focusing and for collecting the PL signal from the sample. A continuous flow cryostat allows for sample cooling down to liquid He temperature. The laser spot diameter on the sample was ~2 μm. The signal is dispersed by a grating spectrometer and recorded by a nitrogen-cooled InGaAs line detector.

**Disk fabrication.** As a demonstrator for the incorporation of GIB-QDs into a resonant cavity we fabricated microdisk-resonators with diameters of 1.8μm. The disk-shape was written by standard electron beam lithography using a negative resist. After development, the resist acts as a mask for a reactive ion etching process in a Oxford 100 cryo-ICP etcher that is used to thin the structures down to the $Si/SiO_2$ interface with perpendicular sidewalls. For this process we used the gases $SF_6$, He and $O_2$. Hereafter, the microcavities are partially underetched by hydrofluoric acid (HF) in order to increase the mode confinement due to the large refractive index contrast between the Si/Ge layer and the surrounding air. The carriers in the sample were optically excited and the PL signal was detected perpendicular to the disk.

The surface roughness of the disk sidewalls and the HF under-etching procedure are not fully optimized, which explains relatively low quality-factors of about 1700 (see supplementary material).

**ACKNOWLEDGEMENTS**

The authors gratefully acknowledge G. Bauer, K. Unterrainer, F. Bechstedt and A. Edwards for helpful discussions, and O. Schmidt and A. Rastelli for support. M.B. and H.G. gratefully acknowledge support from the Austrian Science Fund (FWF): J3328-N19 and J3317-N27. This work was supported by the FWF-funded SFB "IRoN" via grants F2502-N17 and F2512-N17.

## AUTHOR CONTRIBUTIONS

## ADDITIONAL INFORMATION

The authors declare no competing financial interests.

## FIGURE LEGENDS

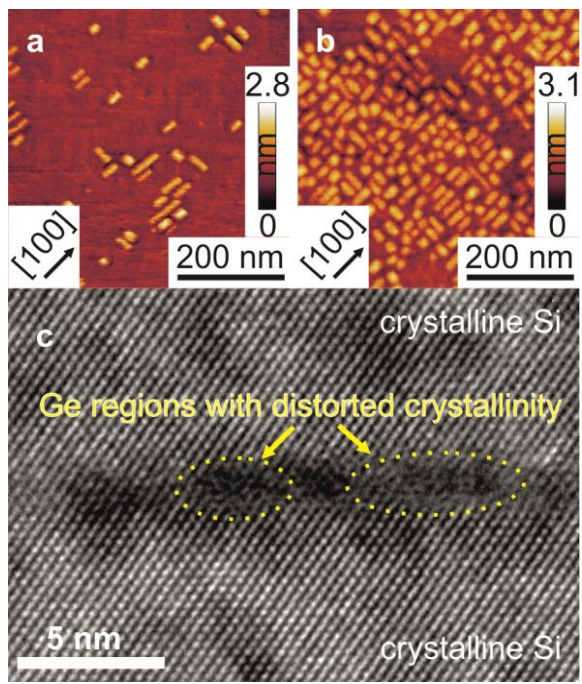

**Figure 1| Structural characterization of GIB-QDs**

**a** and **b** Atomic force microscopy (AFM) micrographs of uncapped GIB-QDs where 0.7 nm and 0.84 nm, respectively, of Ge were deposited at 500°C under $V_{GIB}$ = -2.8 kV. In such a way the GIB-QD density can be tuned from ~$2 \cdot 10^{10}$ cm$^{-2}$ to $2 \cdot 10^{11}$ cm$^{-2}$.

**c** High resolution TEM image of a GIB-QD recorded along the [110]-zone axis. The GIB-QD is embedded in a crystalline Si matrix and partly exhibits a glassy, irregular atomic arrangement due to the partial amorphisation.

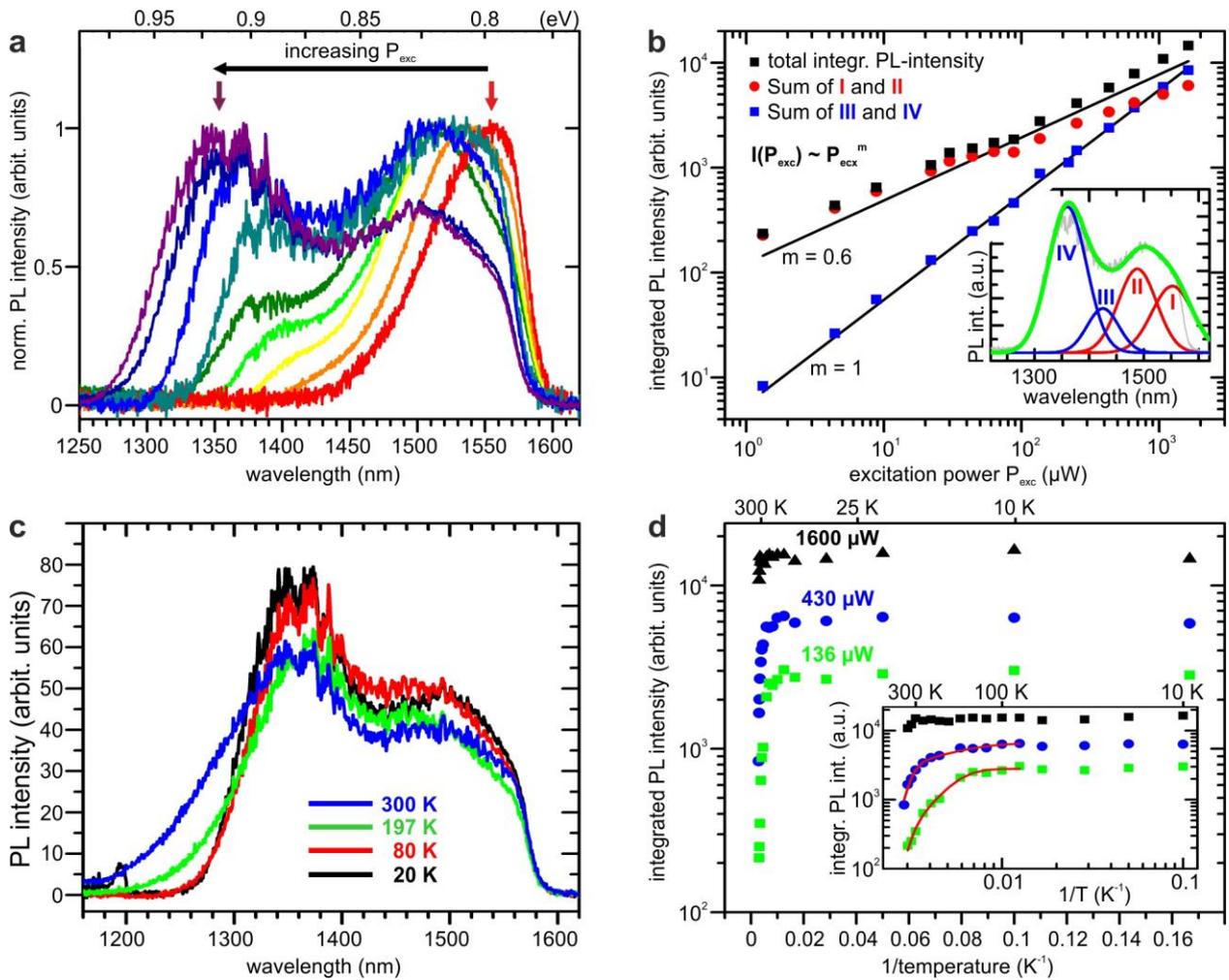

**Figure 2 | Photoluminescence properties of GIB-QDs. a** Normalized PL spectra for increasing $P_{exc}$. The onset of the PL intensity (1450 nm – 1250 nm) shows a blue-shift for increasing $P_{exc}$ due to filling of smaller GIB-QDs that have larger confinement energies and, thus, shorter emission wavelengths. **b** Integrated PL-intensity vs. $P_{exc}$. The PL spectra were fitted with 4 Gaussian functions (see inset). The sum of Gaussians I and II (red-circles), of Gaussians III and IV (blue squares), and the total integrated PL intensity (black squares) are plotted. The black solid lines represent power coefficients $m$ of 0.6 and 1, respectively. **c** PL-spectra of GIB-QDs for $T_{PL}$ of 20 K (black), 80 K (red), 197 K (green) and 300 K (blue). The spectra are not normalized. **d** Integrated PL intensity of the GIB-QDs versus inverse temperature for $P_{exc}$=136 µW (green squares), 430 µW (blue circles) and 1600 µW (black triangles). The inset shows double logarithmic Arrhenius plots. The red curves are fits to the data (see supplementary material), corresponding to $E_A$ of about 350 meV for $P_{exc}$=136 µW and 430 µW. For $P_{exc}$=1.6 mW the PL-quenching at 340K is not pronounced enough to accurately determine $E_A$.

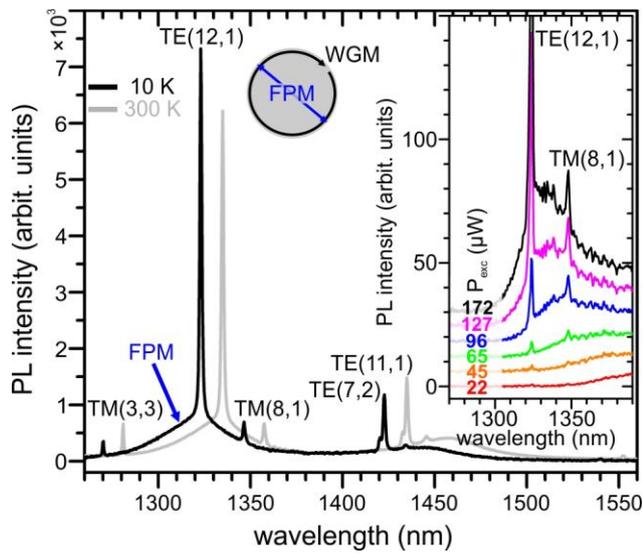

**Figure 3 | PL from of GIB dots in a microdisk.** Mode-PL-spectra are recorded for $T_{PL} = 10$ K (black) and 300 K (grey). The inset schematically depicts the origin of the observed whispering-gallery-modes (WGM) and Fabry-Pérot-modes (FPM). The other inset depicts the $P_{exc}$-dependency of the PL-mode-spectrum. Note the emerging of the transversal electric resonance TE(12,1).

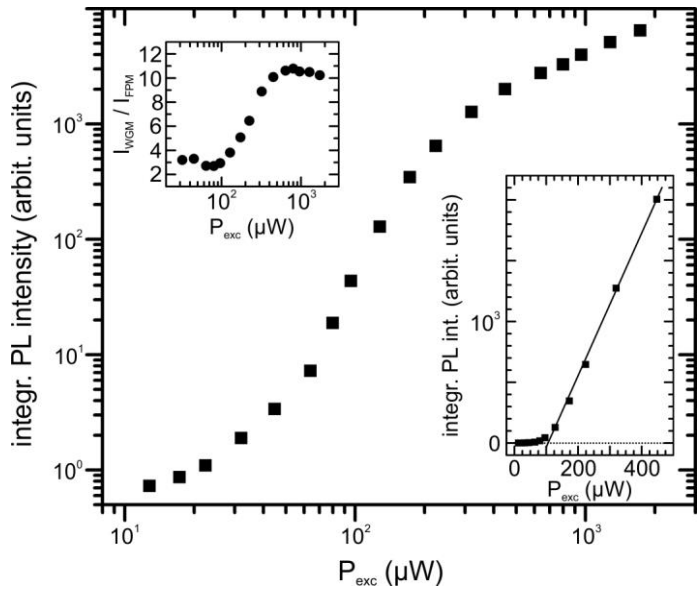

**Figure 4 | $P_{exc}$-dependence of $I_{WGM}$ TE(12,1) on a double logarithmic scale.**

$T_{PL} = 10$ K. The lower inset depicts the threshold-behaviour on a double-linear scale. In the upper inset the ratio $I_{WGM}/I_{FPM}$ at 1323 nm is plotted.

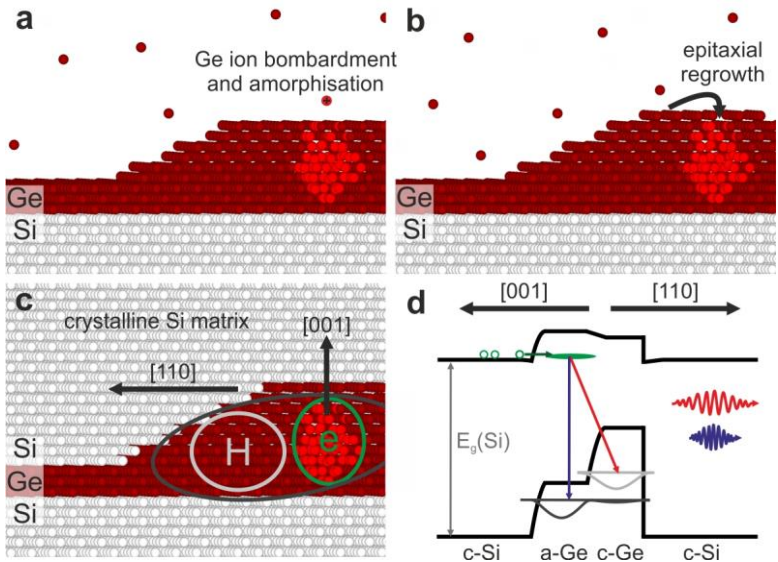

**Figure 5 | Formation of GIB-Ge-dots**

Schematic presentation of the influence of Ge-ion-bombardment (GIB) on a growing GIB-QD. **a** Positively charged Ge ions impinge on the surface causing local, a few nm wide amorphisation of ellipsoidal shape, whereas the rest of the structure keeps its crystalline nature. **b** During the further growth and the ramp-down to the growth-temperature of the Si-capping-layer solid-phase epitaxial recrystallisation takes place reconstructing the top layers of the Ge QD starting from still coherent parts. **c** This layer in turn acts as a base for fully dislocation-free epitaxial overgrowth of the whole GIB-Ge-dots by Si. Small crystalline regions exhibiting 3D confinement are separated by glassy barriers. **d** schematic band-diagram in [001]- and [110]-direction. The green ellipse depicts localized electron states at dangling bonds. The red and blue arrows show GIB-QD-PL emission of longer and shorter wavelength and the hole-wave-functions in the GIB-QD are in grey. The spatial positions of the wave-functions are plotted as light- and dark-grey and green ovals in **c**.

SUPPLEMENTARY MATERIAL

# Efficient room-temperature light-emitters based on partly amorphised Ge quantum dots in crystalline Si

M. Grydlik, F. Hackl, H. Groiss, M. Glaser, A. Halilovic, T. Fromherz, W. Jantsch, F. Schäffler and M. Brehm

Institute of Semiconductor and Solid State Physics, Johannes Kepler University Linz, Altenbergerstrasse 69, A-4040 Linz, Austria

**Content:**

## I. Fits to Arrhenius plots of the GIB-QDs.

In the main text of the manuscript, we determine also activation energies ($E_A$) from fits to Arrhenius plots in order to characterize the thermal quenching of the GIB-QD-related photoluminescence (PL) emission. We consider here a nanostructure-system, and, thus, the proposed transitions involve structures with different dimensionalities. Consequently, the formula to extract the activation energies $E_A$ from the temperature-dependent integrated PL intensity data reads as follows Refs. 1, 2:

$$I_{PL}(T_{PL}) = I_0 \cdot (1 + A \cdot T^{f-i} \cdot \exp(-E_A/k_B T_{PL}))^{-1}, \qquad \text{(Eq. 1)}$$

where $I_0$ is the integrated PL-intensity at 6 K, A is a scaling coefficient, $k_B$ the Boltzmann constant and $T^{f-i}$ results from to the temperature-dependences of the effective density of states in the final state (f) and the initial state (i) (Refs. 1,2). The temperature-dependent pre-factor accounts for the fact that the effective density of states is temperature-dependent with an exponent that depends on the dimensionality.

For our system we consider thermal quenching of carriers located in and around a 0D-QD that thermalize into 3D bulk material. In this case f = 3/2 for bulk Si and i = 0 for the GIB-QDs. Consequently a pre-factor of $T^{3/2}$ has been taken into account in Eq. 1.

## II. Time-resolved-photoluminescence spectroscopy investigations.

For time-resolved measurements covering time scales from several tens of ps up to the μs range we used a μ-PL setup with a microscope objective which focuses the laser beam and collects the PL signal emitted by the sample. The sample was excited by a pulsed laser with a wavelength of 442 nm, a pulse width of less than 200 ps and an average optical power ranging from 7 μW to 440 μW. The time-delayed PL signal was detected by a superconducting single photon detector (SSPD) from Scontel, operated at 1.8 K. It allows single photon detection with a quantum efficiency of approximately 12% (at a wavelength λ = 1310 nm) and a counting rate larger than 70 MHz. For an acquisition of the decay curves, the time-correlated single-photon counting (TCSPC) system PicoHarp 300 by PicoQuant was used which records the arrival times of single photons detected by the SSPD relative to the excitation laser pulse times (start - stop times). The time-resolution of the measurements presented in this work was 64 ps at an excitation pulse rate ranging from 2.5 MHz to 80 MHz. Different emission wavelengths $\lambda_{PL}$ of the PL were selected by band-pass filters with half-width at half-maximum (HWHM) of about 10 nm. Figure S1 presents the location of the bandpass-filters with respect to a GIB-QD-PL spectrum recorded for $P_{exc}=1070\mu W$.

Most of the decay-curves of the GIB-QDs-related PL-emission show neither single- nor double-exponential behavior. They are best described by a log-normal distribution as discussed in detail by Driel et al.[3]. The distribution of decay rates can be extracted following Eq. 2.

$$\sigma(\Gamma) = C \cdot \exp(-((\ln \Gamma - \ln \Gamma_{mf}) /\gamma)^2) \qquad \text{Eq. (2)}$$

Here, C is a normalization constant, γ is related to the HWHM of the distribution, $\Delta\Gamma$ and $\Gamma_{mf}$ is the most frequent rate constant:

$$2\cdot \Delta\Gamma = 2\cdot \Gamma_{mf}\cdot\sinh(\gamma) \qquad \text{Eq. (3)}$$

The average lifetime of the decay is thus calculated as follows:

$$\tau_{av} = 1/\Gamma_{mf} \cdot \exp(\Delta\Gamma^2/2) \qquad \text{Eq. (4)}$$

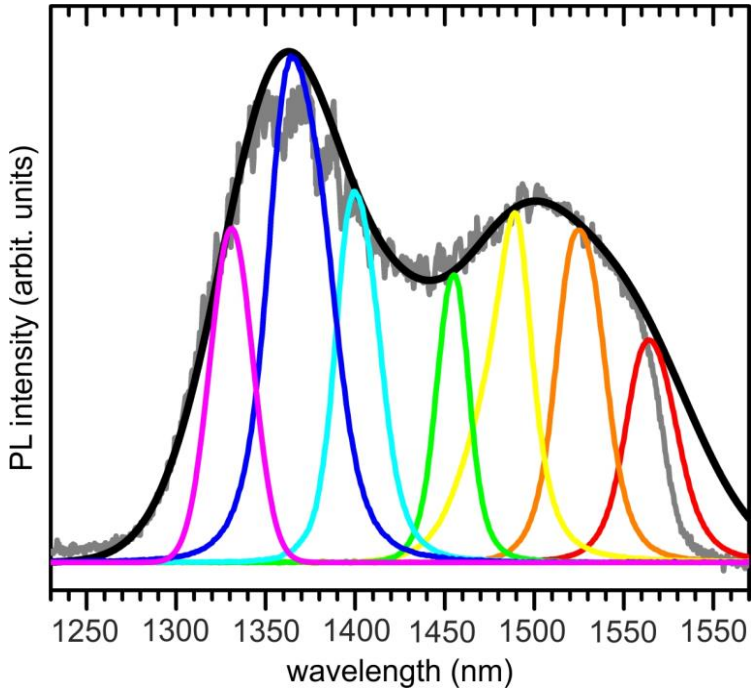

**Figure S1 | GIB-QD-PL spectrum for $P_{exc}$=1070µW.** The wavelengths and widths of the bandpass-filters used for the time-resolved spectroscopy experiments are indicated by red, orange, yellow, green, light-blue, dark-blue and magenta color.

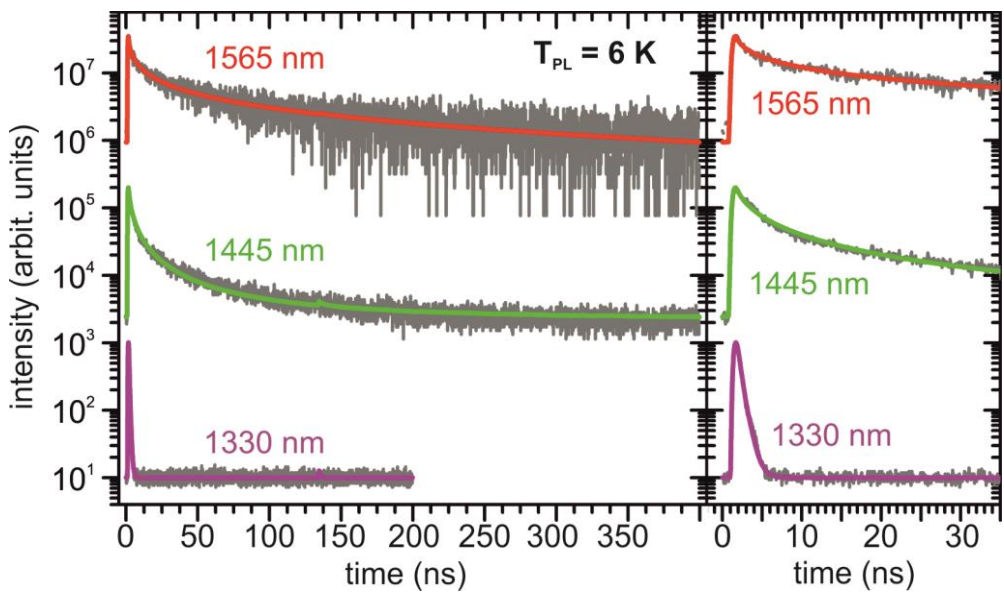

**Figure S2 | Time-resolved PL spectra of GIB-QDs recorded for $T_{PL}$ = 6 K.** The emission was filtered with different band-pass filters, $\lambda_{PL}$ = 1565 nm, 1445 nm and 1330 nm. A distinct difference in the PL decay times is observed. The right panel shows the early response (0 ns – 35 ns) of the decay spectra. The spectra were fitted using a log-normal distribution function. Those fits are shown by the red, green and violet curves.

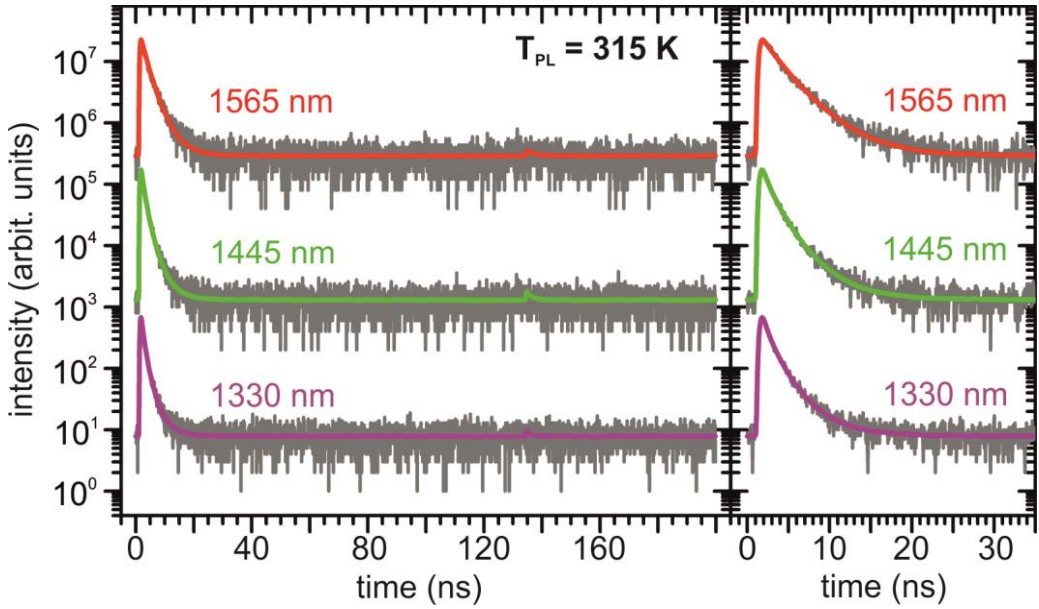

**Figure S3 | Time-resolved PL spectra of GIB-QDs recorded for $T_{PL}$ = 315 K**. The emission was filtered with different band-pass filters, $\lambda_{PL}$ = 1565 nm, 1445 nm and 1330 nm. Independently of the collected PL wavelength a similar average PL decay time of 1-2 ns is observed. The right panel shows the early response (0 ns – 35 ns) of the decay spectra. The spectra were fitted using a log-normal distribution function. Those fits are shown by the red, green and violet curves.

Figure S2 depicts the time-decay of the PL, measured at a sample temperature $T_{PL}$ of 6 K, for the three selected PL-emission wavelengths $\lambda_{PL}$ of 1565 nm, 1445 nm, and 1330 nm. Obviously, the average lifetimes $\tau_{av}$ are distinctively different for the different $\lambda_{PL}$. For $\lambda_{PL}$ = 1565 nm $\tau_{av}$ is about 100 ns, for $\lambda_{PL}$ = 1445 nm about 6 ns and for $\lambda_{PL}$ = 1330 nm about 600 ps. The panel on the right hand side shows the early time decay from 0 to 35 ns. This large difference in the wavelength-dependent average lifetimes completely vanishes at room-temperature and above, as shown in Fig. S3 for $T_{PL}$ = 315 K. $\tau_{av}$ is then between 1and 2 ns, independently of $\lambda_{PL}$.

The whole behavior of the $\lambda_{PL}$- and $T_{PL}$-dependent average lifetimes of the GIB-QD PL is summarized in Fig. S4. For low $T_{PL}$ < 180 K, $\tau_{av}$ is strongly wavelength dependent: it amounts to several hundreds of ns for $\lambda_{PL}$ = 1565 nm and decreases to about 600 ps for $\lambda_{PL}$ = 1330 nm. For high $T_{PL}$ > 300 K $\tau_{av}$ converges to a level of about 1-2 ns, independent of the wavelength. Interestingly, $\tau_{av}$ actually increases to the high-$T_{PL}$-value of 1-2 ns for the short-wavelength PL-contribution.

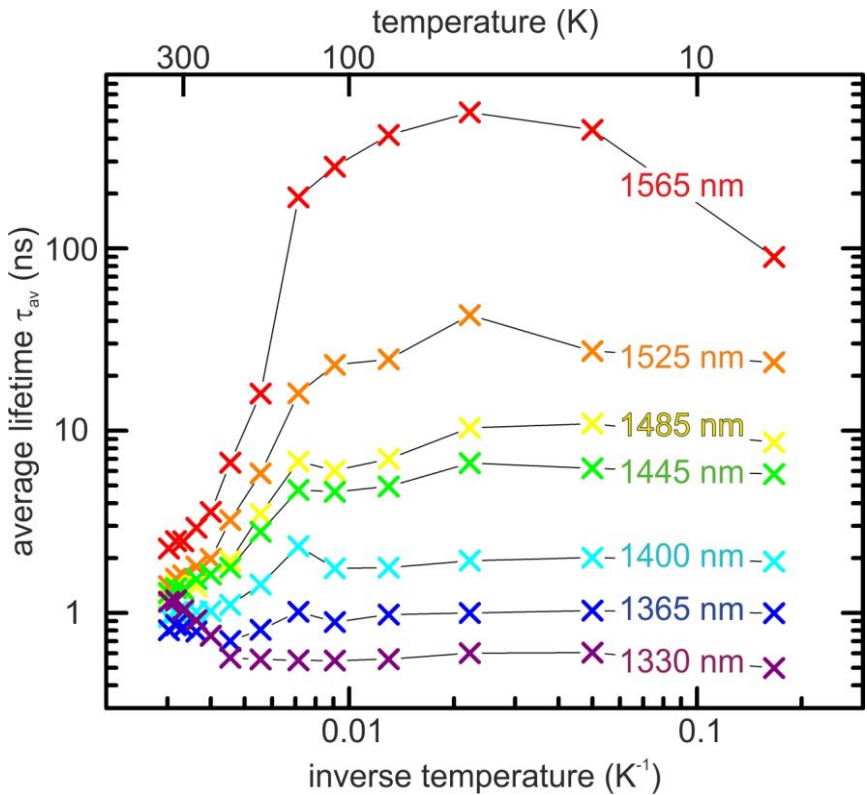

**Figure S4 | T$_{PL}$ dependence of the average lifetimes τ$_{av}$** for different PL λ$_{PL}$ = 1565 nm (red), 1525nm (orange), 1485 nm (yellow), 1445 nm (green), 1400 nm (light-blue), 1365 nm (dark-blue) and 1330 nm (violet). For low T$_{PL}$ < 180 K τ$_{av}$ is strongly λ$_{PL}$-dependent with several hundreds of ns for λ$_{PL}$ = 1565 nm and about 600 ps for λ$_{PL}$ = 1330 nm. For high T$_{PL}$ > 300 K τ$_{av}$ converges to a level of about 1-2 ns, independent of λ$_{PL}$. The black lines are only guides for the eye.

In Figure S5 the logarithm of the half-width at half maximum of the fitted log-normal distribution functions is plotted. This parameter HWHM gives a qualitative measure of the amount of different transition processes participating in the PL time-decay. A logarithmic HWHM of 0 implies a single exponential decay, while higher values for HWHM imply that the curves would have to be fitted with higher numbers of exponential functions. For T$_{PL}$ < 180 K a distinct wavelength dependency of HWHM is observed, while for higher T$_{PL}$ the λ$_{PL}$-dependent values of the HWHM converge.

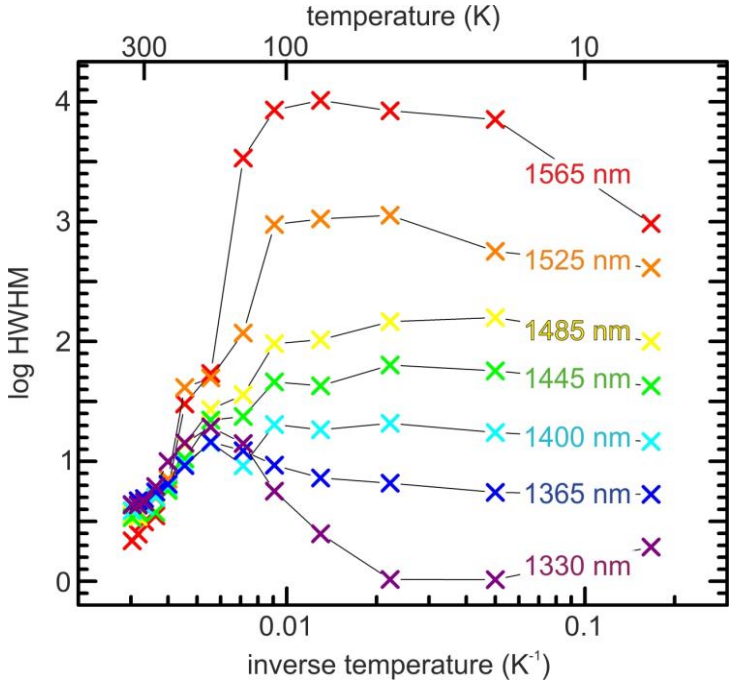

**Figure S5 | $T_{PL}$-dependence of the logarithmic half-width at half maximum** (HWHM) of the lognormal-distribution, used to fit the $\lambda_{PL}$-dependent PL-decay spectra. HWHM gives a qualitative measure of the amount of different processes participating in the PL time-decay. A HWHM of 0 implies a single exponential decay, while higher values for HWHM implies that the curves would have to be fitted with higher numbers of exponential functions. For $T_{PL} < 180$ K a distinct wavelength dependency of HWHM is observed, while for higher $T_{PL}$ the $\lambda_{PL}$-dependent values of the HWHM converge.

Figure S6a depicts the $\lambda_{PL}$-dependent integrated PL-intensity as obtained from the time-resolved PL-spectroscopy experiments. For low $P_{exc}$, $I_{PL}$ remains constant while for high $P_{exc}$ the PL-intensity quenches, in agreement to the results reported in Fig. 2d of the main text of the manuscript. In Fig. S5b we additionally plotted the evolution of $I_{PL}$ at high temperatures ($T_{PL}$>77 K). For $\lambda_{PL} =$ 1330nm, $I_{PL}$ actually increases for 140 K < $T_{PL}$ < 275 K, while for $T_{PL}$ > 275 K $I_{PL}$ starts to thermally quench.

Taking into account the results of Figs. S4, S5 and S6 in combination with the results of Fig. 2 of the main text, we conclude that for low $T_{PL}$ (<180K) the PL-decay lifetimes are mainly dominated by radiative transitions, while for higher $T_{PL}$ thermal quenching and non-radiative recombination processes take over. Those non-radiative recombination processes exhibit almost exponential PL time-decay behavior (Fig. S5) and exhibit a $\tau_{av}$ of about 1-2 ns, independently of the PL-emission wavelength. For smaller $\lambda_{PL}$, i.e. for the smaller dots, we observe an increase in $\tau_{av}$ in the range of 140 K < $T_{PL}$ < 275 K before at higher $T_{PL}$ thermal quenching of $I_{PL}$ sets in. It follows from the increase of $I_{PL}$ with increasing $T_{PL}$, that the GIB-QD system cannot be regarded as a simple two energy-level system, but that additionally trap-states, possibly originating from trivalent bonds of the crystalline-glassy-interface or refilling of carriers from different – larger – GIB-QDs has to be taken into account. In the future, further investigations of the carrier dynamics of the novel GIB-QDs will have to be performed to gain more insight into the contributing recombination mechanisms.

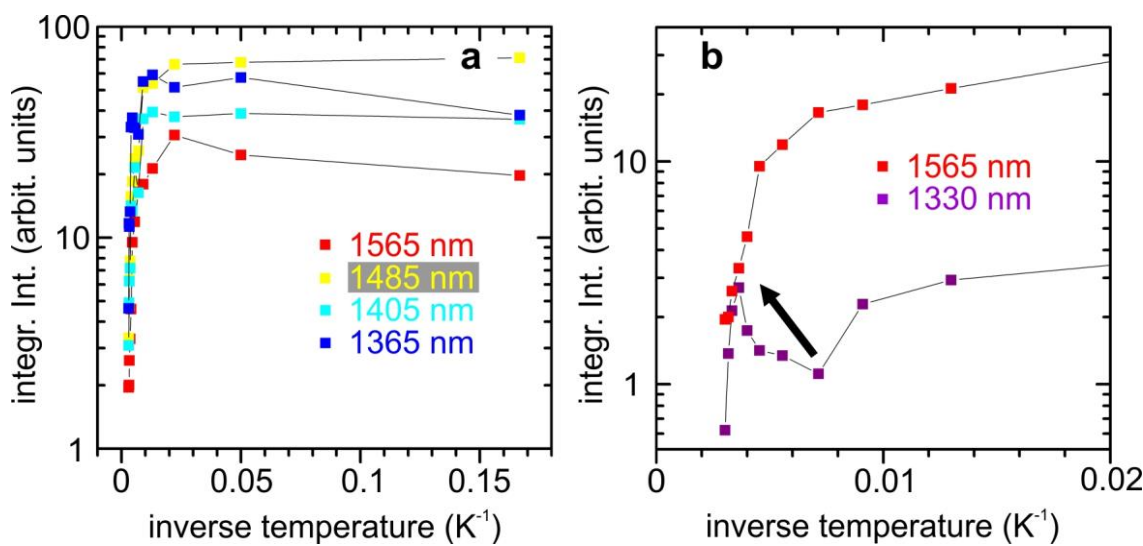

**Figure S6 | $I_{PL}$ of the time-resolved PL-spectroscopy experiments versus $T_{PL}$. a** $I_{PL}$ vs $T_{PL}$ for $\lambda_{PL}$ of 1565 nm (red), 1485 nm (yellow), 1405 nm (light green) and 1365 nm (dark blue). **b** $I_{PL}$ vs $T_{PL}$ for $\lambda_{PL}$ of 1565 nm (red) and 1330 nm (violet). Note the increase of $I_{PL}$ for 140K < $T_{PL}$ < 275 K for $\lambda_{PL}$ =1330nm. A similar but less pronounced increase of $I_{PL}$ for this $T_{PL}$ window is observed for $\lambda_{PL}$ =1365nm (**a**).

## III. TEM investigations of the wetting layer.

In the main text of the manuscript we stated that only the about 3.5 nm high GIB-QDs are influenced by amorphisation by low energy Ge ion bombardment, whereas the WL remains unaffected. This is depicted in Fig. S7 where a high-resolution transmission electron microscopy (TEM) image of the WL between the GIB-QDs ($V_{GIB}$ = -2.8 kV) is presented. The Si matrix material as well as the WL are fully crystalline and do not show signs of amorphisation or other defective behaviour. Here we argue that the absence of amorphisation can be explained by the small thickness of the wetting layer of about 4 monolayers (ML). Even if the WL is getting amorphised during Ge ion bombardment, such a thin layer will recrystallize during growth[4,5].

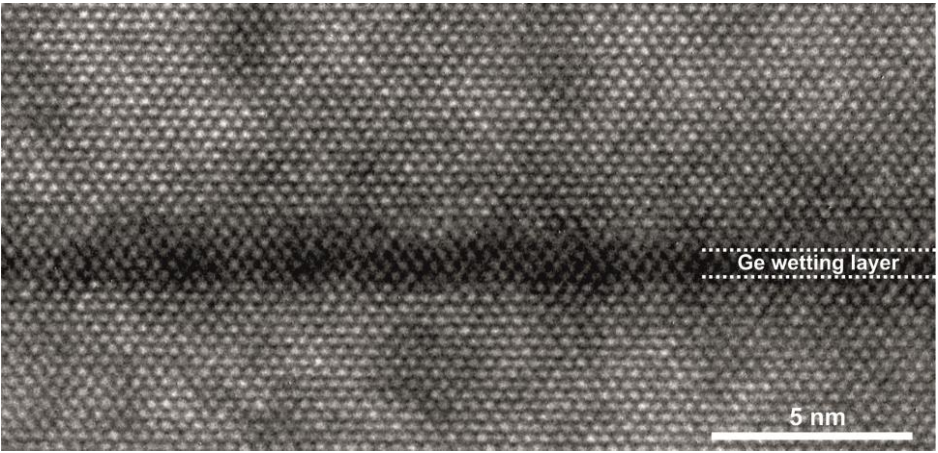

**Figure S7 | High-resolution TEM image of a wetting layer between the GIB-QD** ($V_{GIB}$ = -2.8kV) recorded along the [110] zone axis. The fully crystalline WL, embedded in a crystalline Si matrix, does not show any signs of amorphisation.

## IV. Dependence of the acceleration voltage $V_{GIB}$ on the PL-emission of the GIB-QDs.

We studied the influence of the Ge-ion bombardment energy on the volume of amorphised regions as traced by PL-spectroscopy. With increased $V_{GIB}$ and thus increased volume of the amorphised zones we expect the remaining crystalline regions in the GIB-QDs to shrink. We prepared a set of samples where $V_{GIB}$ was altered as the only growth parameter during the Ge layer growth. The growth conditions were kept constant and the active region of 6 ML of Ge grown at 500°C is the same for all the samples. All samples were overgrown with a 100 nm thick Si capping layer deposited at 350°C.

In Figure S8 we present low-temperature (10 K) PL-spectra of GIB-QDs, for which $V_{GIB}$ was -0.4, -1.3, -2.0 and -2.8kV, respectively. The relatively high noise of the spectra (grey color) is related to the sensitivity of the wavelength-extended InGaAs line detector used for this experiment. The samples were located in a bath cryostat and excited with an $Ar^+$ laser operating in cw-mode at 514.4 nm with $P_{exc}$ = 5 mW and a spot diameter of about 300 μm. The red curves represent smoothened spectra and are a guidance for the eyes. The growth rate was 0.15 Å/s for all the samples. The topmost spectrum is from a sample with $V_{GIB}$ =-2.8 kV, grown however at a rate of 0.015 Å/s which leads to a higher effective dose of Ge ions for this sample. The intensity maximum of the GIB-QD peaks is at 0.73 eV for $V_{GIB}$ = -0.4 kV. For higher $V_{GIB}$ of -1.3 kV, -2.0 kV and -2.8kV the GIB-QD-related PL peaks are blue-shifted to 0.75 eV and 0.77eV, respectively. For the sample with $V_{GIB}$ = -2.8 kV where Ge was deposited at 0.015 Å/s the GIB-QD-PL peak is even further blue-shifted to 0.93 eV.

The blue-shift of the GIB-QD's PL signal with increasing $V_{GIB}$ and increasing effective Ge-ion dose can be explained by the decreasing size of the crystalline Ge regions and the concomitant increase of the quantum confinement energy in those parts of the GIB-QDs. The glassy, zones within the Ge islands are enlarged as $V_{GIB}$ is increased. Summarizing, we conclude from this experiment that the holes very likely are located in the crystalline parts of the GIB-QDs.

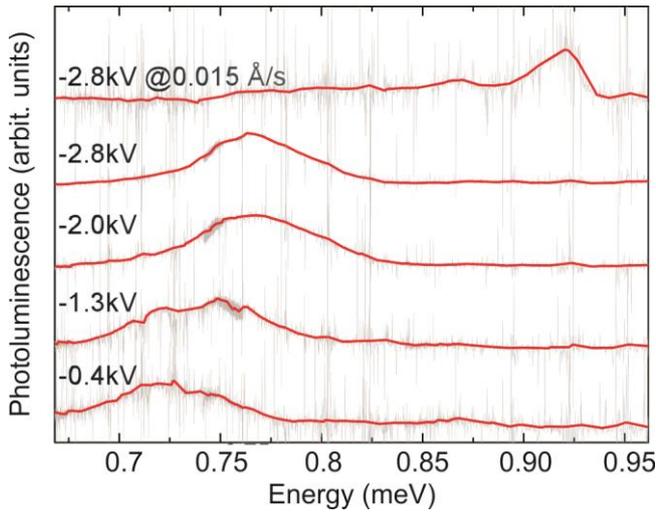

**Figue S8| Photoluminescence intensity of GIB-QD samples**, where $V_{GIB}$ was varied from -0.4kV (lowest spectrum) to -2.8kV (topmost). The Ge growth rate was 0.15 Å/s for all samples but the topmost one, for which the rate was lowered to 0.015Å/s, resulting in a higher effective dose of Ge ions bombarding the QDs.

## V. Additional data concerning PL-properties of microdisk containing GIB-QDs.

Figure S9a shows the $P_{exc}$-dependent intensity ratio of the whispering gallery mode TE(12,1) as compared to the Fabry-Perot mode, emitting at the same wavelength, given also in the inset of Fig. 4 of the main manuscript. In Figure S9b and c the Q-factor and the full width at half maximum FWHM of the Lorentzian-shaped peak TE(12,1) are presented. The Q-factor is determined by the ratio of the emission wavelength of the peak versus FWHM. At the observed threshold of $I_{PL}$ and also $I_{WGM}/I_{FPM}$ we also observe a slight increase of the Q-factor from about 1400 to about 1700. This increase of the Q-factor at the threshold is a direct consequence of the slight linewidth-narrowing of the WGM.

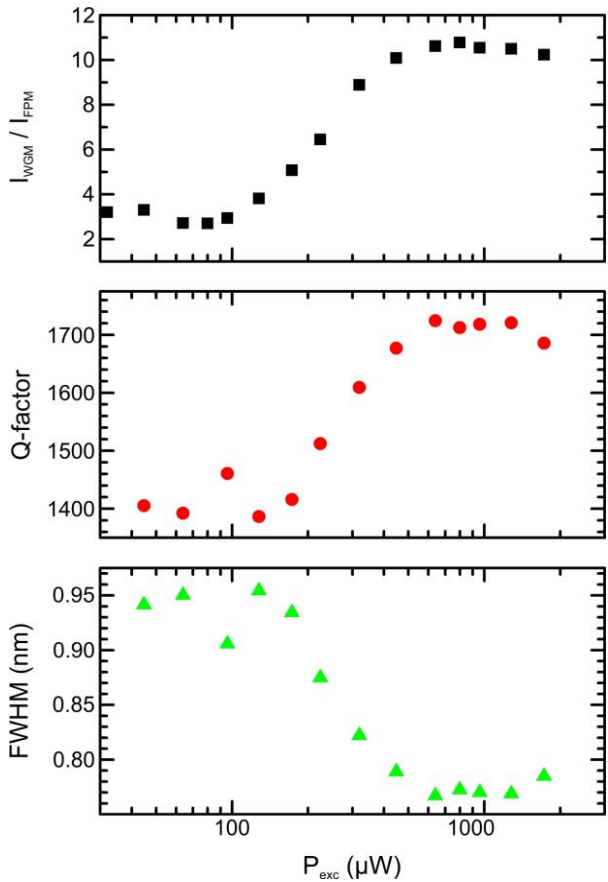

**Figure S9 | a Ratio I_WGM / I_FPM, b Q-factor and c FWHM of WGM resonance TE(12,1) versus P_exc.** $T_{PL} = 10K$. Data recorded for cw-laser excitation. Concomitant to the increase of IWGM /IFPM versus $P_{exc}$ also the Q-factor of the observed resonance TE(12,1) increases while the FWHM of the resonance decreases.

Threshold behaviour of the PL-intensity, in combination with linewidth-narrowing and decrease of the average PL-decay-lifetimes suggests stimulated emission of the GIB-QDs in the photonic resonator.

## VI. References.